\documentclass[reviewcopy]{elsarticle}

\usepackage[reviewcopy]{adndt}
\usepackage{longtable}

\usepackage{amsmath}

\usepackage{amssymb}

\biboptions{square,sort&compress}
\bibpunct[]{[}{]}{,}{n}{}{;}
\begin{document}

\begin{frontmatter}

\journal{Atomic Data and Nuclear Data Tables}


\title{Energy levels and radiative rates  for transitions in Cr-like Co IV and Ni V}

  \author[One]{K. M. Aggarwal\corref{cor1}}
  \ead{K.Aggarwal@qub.ac.uk}
 \author[Two]{P. Bogdanovich\fnref{}} 
 \author[Two]{R. Karpu\v{s}kien\.e\fnref{}} 
  \author[One]{F. P. Keenan\fnref{}}
  \author[Two]{R. Kisielius\fnref{}} 
   \author[Three]{ V. Stancalie\fnref{}} 


  \cortext[cor1]{Corresponding author.}

  \address[One]{Astrophysics Research Centre, School of Mathematics and Physics, Queen's University Belfast,\\Belfast BT7 1NN,
Northern Ireland, UK}

  \address[Two]{Institute of Theoretical Physics and Astronomy, Vilnius University, A. Go\v{s}tauto 12, LT-01108 Vilnius, Lithuania}
  \address[Three]{National Institute for Laser, Plasma and Radiation Physics, Atomistilor 409, P.O. Box MG-36, Magurele-Ilfov, 077125 Romania}
\date{16.12.2002} 

\begin{abstract}  
We report calculations of energy levels and radiative rates  ($A$-values) for transitions in Cr-like Co IV and Ni V.  The quasi-relativistic Hartree-Fock  (QRHF) code is adopted for calculating the data although {\sc grasp} (general-purpose relativistic atomic structure package) and flexible atomic code ({\sc fac}) have also been employed for  comparison purposes. No radiative rates are available in the literature  to compare with our results, but our calculated energies are in close agreement  with those compiled by NIST for a majority of the levels. However,  there are discrepancies for a few levels of up to 3\%. The $A$-values are listed for all significantly contributing E1, E2 and M1  transitions, and the corresponding  lifetimes  reported, although unfortunately no previous theoretical or experimental results exist  to compare with our data.\\  \vspace*{0.5 cm}

{\em Received}: 14 May 2015, {\em Accepted}: 24 September 2015 


\end{abstract}

\end{frontmatter}




\newpage

\tableofcontents
\listofDtables
\listofDfigures
\vskip5pc


\section{Introduction}

 Iron group elements (Sc -- Zn) are becoming increasingly important in the study of astrophysical plasmas, as many of their  lines from different ionization stages  are frequently observed. These lines provide a wealth of data on plasma characteristics, including  temperature, density and chemical composition. Additionally, iron group elements are often impurities in fusion reactors, and to estimate the power loss from the impurities, atomic data (including energy levels and  oscillator strengths or radiative decay rates) are required for many ions. The need for atomic data has become even greater with the developing  ITER project. Since there is a paucity of measured parameters, one must depend on theoretical results. Therefore,  over the last few years  we have reported atomic parameters for many ions of the iron group elements -- see for example  \cite{fe26,ni19,kr35,co25} and references therein. Among Co ions, results have already been provided for Co XXVII \cite{co27}, Co XXVI \cite{co26}, Co XXV \cite{co25}, Co XXII \cite{co22}, Co XVI \cite{co16}, and Co XI \cite{co11}. Similar data for several Ni ions have also been reported, i.e.  Ni XI  \cite{ni11}, Ni~XIII -- Ni~XVI \cite{niions}, Ni~XVII \cite{co16}, Ni~XIX \cite{ni19}, Ni~XXIII \cite{co22}, Ni~XXVI \cite{co25}, Ni~XXVII \cite{co26}, and Ni~XXVIII \cite{co27}. Here we focus our attention on Cr-like Co IV and Ni~V.

Several emission lines of Co and Ni ions have been observed in astrophysical plasmas, as listed in the CHIANTI database at  ${\tt {\verb+http://www.chiantidatabase.org+}}$. Similarly, many lines of these ions, below 2000 $\rm \AA$, are also listed in the compilation by Kelly \cite{kelly}. However, we are not aware of any observed lines of Co IV but Raassen and Hansen \cite{ni5} have identified forbidden lines of Ni~V in Eta Carinae ($\eta$ Car). Similarly, Preval et al. \cite{ong2} have analysed many lines of Ni~V in the near- and far-ultraviolet range (1150--3145 {\rm\AA} and 910--1185 {\rm\AA}) belonging to the 4s--4p transitions, from the high-resolution spectra from hydrogen-rich (DA) white-dwarf star G191-B2B, obtained from the Hubble Space Telescope Imaging Spectrograph (STIS), and determined photospheric abundance of many elements, including Ni. Furthermore, many  emission lines of Co IV and Ni~V are listed in the 200-475 \AA ~wavelength range in the {\em Atomic Line List} (v2.04) of Peter van Hoof at ${\tt {\verb+http://www.pa.uky.edu/~peter/atomic/+}}$, because these are useful in the generation of synthetic spectra. Similarly, experimental data are available in the literature for Co IV and Ni~V  by Poppe  et al.~\cite{pop}, who identified the multiplets of 3d$^6$ $^5$D -- 3d$^5$($^6$S)4p $^5$P$^o$ lines from a laboratory spectrograph. Following these identifications, they calculated energies for several other levels of the (3d$^5$) 4s and 4p configurations, based on some analytical expressions. These energies were compiled by Sugar and Corliss~\cite{sc} and are available on the website of NIST (National Institute of Standards and Technology) at {\tt http://www.nist.gov/pml/data/asd.cfm}  \cite{nist15}. 

The situation with regard to  theoretical results is similar. Energies for a few levels of the (3d$^5$) 4s and 4p configurations are available \cite{ams1,ams2}, from calculations based on {\em least square fitting} of Slater-Condon parameters, and are biased towards the known (observed or measured) results and iso-ionic, iso-electronic and iso-nuclear trends. A wider set of data for  136 terms of the 3d$^6$, 3d$^5$4s and 3d$^5$4p configurations has been reported for Co~IV by  one of the current authors  \cite{vs1},  calculated in $LS$ coupling with the CIV3 code \cite{civ3}, which neglects two-body relativistic operators, although these are not very important  for this moderately heavy ion. More importantly,  limited CI (configuration interaction) with some configurations involving the 4s, 4p and 4d orbitals is  included, and $A$-values are not reported. Similarly, Kingston et al. \cite{ni5a} calculated energies but  for only a few $LS$ states of Ni~V. Therefore, the available data  are not suitable for applications in plasma modelling, because results are required for {\em fine-structure} levels and their corresponding transitions. However, recently Ong et al. \cite{ong1} have reported energies for 131 levels of Ni~V. They adopted the methodology of {\em configuration interaction}  and {\em many-body perturbation theory} (CI+MBPT) for their calculations and the energies obtained agree within 2\% of the NIST compilation.  However, their reported levels are only a subset of the NIST compilation and particularly {\em missing} are the lowest 34 levels of the 3d$^6$ configuration. Additionally, they have not calculated the $A$-values. Therefore, there is scope for extending their calculations.  

In this work we report atomic data for energy levels and $A$-values for transitions among the 3d$^6$,  3d$^5$4s and 3d$^5$4p configurations  of Co IV and Ni~V.  For the calculations, we first  employed the fully relativistic {\sc grasp} (general-purpose relativistic atomic structure  package) code  originally developed by Grant  et al.~\cite{grasp0}, but later revised by one of its authors (P.H. Norrington).  The code is  based on the $jj$ coupling scheme, is referred to as GRASP0 and is available at the website  
{\tt http://web.am.qub.ac.uk/DARC/}.  Furthermore, it includes the  major modifications made in other  versions, such as {\sc grasp92} \cite{grasp2} and {\sc grasp2k} \cite{grasp2k,grasp2kk}. Relativistic corrections arising from the Breit interaction and QED  (quantum electrodynamics) effects (vacuum polarization and Lamb shift) are also  included. Finally, we have used the option of {\em extended average level} (EAL),  in which a weighted (proportional to the level statistical weight $2j+1$) trace of the Hamiltonian matrix is minimized. This produces a compromise set of orbitals describing closely lying states with  moderate accuracy. However, it soon became clear that the level of CI required for Cr-like ions is too large, and the desired calculations  could not be performed with the {\sc  grasp} code within a reasonable time frame (a few months). Since our  aim is to calculate only energy levels and $A$-values, we hence adopted the  {\em Flexible Atomic Code} ({\sc fac}) of Gu~\cite{fac}, available from the website 
{\tt {\verb+https://www-amdis.iaea.org/FAC/+}}. This is also a fully relativistic code  and  yields results for energy levels and radiative rates comparable to {\sc grasp}, as has already been noted in several of our earlier papers  -- see, for example, Aggarwal  et al.~ \cite{co16}.  The {\sc fac}  code is comparatively more  efficient and therefore  several sets of calculations with varying amount of CI can be performed within a reasonable time frame without significant  loss of accuracy. However, in spite of including very large CI (see section 2), differences in energies compared to the NIST compilations remained unacceptably large (up to 35\%) for many levels. This motivated us to try yet another approach with much more extensive CI, possible with our quasi-relativistic  Hartree-Fock (QRHF) method that loosely follows ideas  originally developed by Cowan \cite{rdc}. The presently-adopted code has been developed by Bogdanovich and Rancova \cite{qr06,qr07} and includes packages from    \cite{ah91, cff91, cffmg91}. Its  main properties  are
summarized by Bogdanovich and Kisielius \cite{bk12} and it has been successfully employed for a variety of ions -- for example, highly ionized W  \cite{bk12} and singly ionized S ~\cite{s2}.

\section{Energy levels}

As noted above, we first employed the {\sc grasp} code to calculate energy levels and $A$-values for transitions in Co~IV. We performed a series of calculations with increasing CI with up to 21 configurations, namely 3d$^6$,  3d$^5$4$\ell$, 3d$^5$5$\ell$, 3d$^4$4$\ell^2$, 3d$^4$4$\ell$4$\ell'$, and 3p$^5$3d$^7$. These configurations  generate up to 14~732 levels, and for brevity we focus on results only from the  final calculations with the {\sc grasp} code, which are listed in Table~A (including the Breit and QED corrections) for the lowest 33  levels of the 3d$^6$ and  3d$^5$4s  configurations. We  note  that the Breit contributions are up to 0.004 Ryd whereas those from QED are negligible. Also included in this table for comparison are the results compiled by NIST. It is clear from the table that there is a significant discrepancy, for a majority of levels, between our results with {\sc grasp} and those of NIST -- see for example, levels 6, 24 and  33, for which the differences are up to 26\%. For most levels our results are lower, but are higher for a few, such as 6--32. Furthermore, the orderings are different between the two sets of energies -- see for example, levels 6--14. Additionally, there are similar differences  with our other calculations with lesser CI, which are not included in Table~A for brevity. One reason for the large discrepancies for many levels may be the inadequate inclusion of CI in the generation of our wavefunctions. Therefore, to explore this limitation we have undertaken a series of calculations with the {\sc fac} code, which we discuss below.

We have performed several sets of calculations with differing amount of CI, specifically the following five: (i) FAC1: 1334 levels among the 3d$^6$,  3d$^5$4$\ell$, 3p$^5$3d$^7$, 3p$^4$3d$^8$, 3p$^3$3d$^9$, and 3p$^2$3d$^{10}$ configurations, (ii) FAC2:  21~992 levels of the 3d$^6$,  3d$^5$4$\ell$, 3d$^5$5$\ell$, 3p$^6$3d$^4$4$\ell$4$\ell'$, 3p$^5$3d$^7$, and 3p$^4$3d$^7$4$\ell$ configurations, (iii) FAC3: including an additional 37~198 (total 59~190) levels of the 3p$^4$3d$^8$, 3p$^3$3d$^9$, 3p$^2$3d$^{10}$, 3p$^6$3d$^4$5$\ell$5$\ell'$, and 3p$^4$3d$^7$5$\ell$ configurations, (iv) FAC4:  including a further 12~338 (total 71~528) levels of the 3p$^5$3d$^6$4$\ell$ and 3p$^5$3d$^6$5$\ell$  configurations, and finally (v) FAC5: 76~138 levels which include the additional 4610 levels of the 3p$^3$3d$^8$4$\ell$,  3p$^2$3d$^9$4$\ell$ and 3p3d$^{10}$4$\ell$ configurations. Further inclusion of additional CI was not considered because the energies have now {\em converged}, i.e. the differences between the FAC4 and FAC5 energies and their level orderings are negligible. Therefore, energies obtained with our final FAC5 calculations/configurations are also provided in Table~A.

Unfortunately, significant discrepancies with the NIST listings remain for a majority of the levels, in both energies (up to 35\%) and their orderings - see for example, level 33, i.e. 3d$^5$($^6$S)4s $^7$S$_3$. Although the NIST energies are based on measurements of only 9 levels among the lines of the 3d$^6$ $^5$D -- 3d$^5$($^6$S)4p $^5$P$^o$ multiplet, the discrepancies between theory and experiment are too large. Moreover, no systematic improvement over the theoretical energies is observed, as may be seen in Table~A. For most levels the FAC energies became closer to those of NIST (see levels 2/3 and 10/11) but became worse for others, particularly for 33. These large discrepancies  have prompted us to try a different approach with much more extensive CI than considered so far.  This is because the different approach has been highly successful for some ions in reproducing theoretical energies closer to measurements, as demonstrated by J\"{o}nsson  et al.~\cite{jgg} for B-like O IV and Froese Fischer \cite{cff} for Br-like W XL.  For example, J\"{o}nsson  et al. ~included over 800~000 and 1~000~000 CSFs (configuration state functions) for the even and odd states of O IV to determine the $n \le 3$ energy levels to within 0.25\% of the measurements. Additionally, the necessity of including a much more extensive CI became apparent with the recent calculations of Ong et al. \cite{ong1} for Ni~V. They considered up to $n$ = 12 ($\ell \le$ 3) configurations for their calculations, although concluded that up to $n$ = 5 configurations produce comparatively more accurate results.

In our QRHF calculations of energy for the levels of the 3d$^6$, 3d$^5$4s and 3d$^5$4p configurations, radial orbitals (RO) for the adjusted configuration electrons are determined by the way of solutions of quasi-relativistic Hartree-Fock equations \cite{qr06,qr07}. For the CI wavefunction expansion, we supplement this RO basis with the transformed radial orbitals (TRO) with principal quantum number $5 \leq n \leq 8$ and all possible values of the orbital quantum number $\ell$. These TRO are constructed in a way to ensure the maximum of correlation corrections \cite{tro99,tro08}. We use the same basis of the orthogonal RO both for the even and the odd configurations. This helps us to overcome the issue of non-orthogonality when line strengths for the electric dipole and higher-multipole-order transitions between the configurations of different parity are calculated. 

Unfortunately, the chosen type of RO requires a considerably large CI basis when calculating the energy spectra. The relativistic effects are included in the Breit-Pauli approximation adapted for the quasi-relativistic RO, see Bogdanovich and Rancova \cite{tro08}. We consider the virtual one- and two-electron excitations from all $3\ell ^N$ and $4\ell$ shells of the investigated configurations. This leads to more than 1000 configurations, both the even and the odd ones in the basis of the determined RO. For these configurations, the number of CSFs exceed $10^9$. To reduce the size of the calculations, we select only those admixed configurations that have the largest contributions to the CI wavefunction of the adjusted configuration. This contribution is assessed in the second order of many-body perturbation theory \cite{mbpt05}.  Following this method, we have selected 568 even and 310 odd configurations, and the  corresponding numbers of CSF for these  are 4~110~279 and 7~412~922, respectively. After the CSF-reduction procedure (for details see Bogdanovich  et al.~\cite{csf02}), the corresponding numbers are 656~832 and 907~014. These CSF bases are sufficient  for  further calculations, and have been employed to determine the Co IV and Ni~V data presented here. 

The energies obtained with the QRHF calculations are also listed in Table~A for Co~IV, and it is clear that these  are much closer to the  NIST values. The maximum discrepancy between theoretical and experimental energies for any level is  3\%, which is highly satisfactory considering the complexity of the Co IV ion. In fact, energies obtained for most levels higher than those listed in Table~A are comparatively in better agreement with those of NIST. This may be seen from Table~1, in which energies for the lowest 322 levels belonging to the 3d$^6$, 3d$^5$4s and 3d$^5$4p configurations of Co IV are listed. The percentage differences with the NIST energies are less than 1\% for the excited configurations 3d$^5$4s and 3d$^5$4p,    and less than 3\% for the ground configuration 3d$^6$. In this table we also list the percentage difference with the NIST energies, and note that the mean of the absolute values  for all levels given in the NIST database is only 0.60\%. Finally, we note that the $LSJ$ designations listed in the table  are not definitive and are only for guidance. This is because many levels are affected by configuration mixing and hence their description using just a $LSJ$ notation cannot be unambiguous in all cases. All such levels are shown in {\bf bold faces}.  This situation arises as our code assigns level notation by the maximum percentage contribution for the particular level CI wavefunction expansion, leading to some levels having the same identification. For example, see levels 121 and 126 which have the same CSF designation, i.e.  3d$^5(^4_5$G)4p\,$^5$F$_3^{\mathrm{o}}$. A similar situation arises for the pairs of levels 154/158, 192/213, 214/217, 221/252, 234/255, and 236/256.
The NIST database employes double-level notation in such cases. This is a common problem among all calculations, particularly for those ions whose levels highly intermix, such as Co IV and Co XI \cite{co11}. 

NIST energy orderings are different only in a few instances, such as for levels 11/12 and 39--41, and the discrepancies in energies  between these levels are very small. The configurations investigated  considered in the present work (3d$^6$, 3d$^5$4s and 3d$^5$4p) produce 322 energy levels in total, while the NIST database gives only the  energies  for 296 $LSJ$ levels. That means that we have predicted energies for 26 levels and determined ab initio values for the complete set of the configurations 3d$^6$, 3d$^5$4s and 3d$^5$4p. As the discrepancies between our results and those of NIST are  generally less than 0.5\%, we can safely assume this to be the accuracy of our calculations for the additional 26 levels.

Energies for a few levels (missing from the NIST listings) of the 3d$^5$4s and 3d$^5$4p configurations have been predicted by van het Hof   et al.~\cite{ams1} and Uylings   et al. ~\cite{ams2}. Their calculations are based on a least-square fitting of Slater-Condon parameters for the measured results for other levels. In Table~B we compare our data with their predictions for Co~IV. Generally, our energies for the  3d$^5$4s levels are higher whereas for 3d$^5$4p are lower. However, all differences are less than 1\% and hence there is no significant discrepancy. Furthermore, we note that the calculated energies of \cite{ams1} for the 3d$^5$4s $^5$D levels differ between --5.2\% and 3.9\% with those of NIST (see their Table  XII), whereas our {\em ab initio} energies agree within 3\% for all levels. Therefore, based on this comparison and that  in Table~A, we can confidently state that our calculated energies for all levels of Co IV should be accurate to better than 3\%. 

Our calculated energies for the levels of Ni~V are listed in Table~2 along with those of NIST and Ong et al. \cite{ong1}. NIST compilations are available for most of the levels and differences with our calculations are below 3\%.  As for Co~IV, significant discrepancies in magnitudes are only for the levels of the 3d$^6$ configuration (1--34), and for the  higher ones the agreement is actually better than 1\%. Similarly, the energies of Ong et al., available for only 131 levels, also agree within 2\% with the NIST compilations -- see for example levels 35--36, 99--106, and particularly 180--182, i.e. the level orderings of Ong et al. {\cite{ong1} also differ with those of NIST in a few instances. Nevertheless,  for the {\em common} levels our calculations are comparatively closer to those of NIST, in both magnitudes and orderings. This may be partly due to the adoption of different approaches  in the calculations, and partly due to the comparatively larger CI included in our work. Anyway, the noted discrepancies are still minor. Finally,  in Table~C we compare our energies with those of van het Hof   et al.~\cite{ams1} and Uylings   et al. ~\cite{ams2}, possible for only a few levels. The (dis)agreements between these and our calculations are similar to those for the levels of Co~IV, seen in Table~B.

\section{Radiative rates}

To  our knowledge, no $A$-values are available in the literature for transitions in Co~IV and Ni~V. Therefore, in Table~3  we list energies/wavelengths ($\lambda$, \AA), radiative rates ($A$-values, s$^{-1}$),  oscillator strengths (f, dimensionless), and transition line strengths ($S$-values in atomic unit) for the E1 (electric dipole)  transitions of Co~IV from the lowest 36 (which include all levels of the 3d$^6$ ground configuration -- see Table 1) to higher excited levels. Similar results for  Ni~V are listed in Table~4 and full tables are available online in the electronic version.  For the E2 (electric quadrupole)  and M1 (magnetic dipole) transitions only $A$-values are listed, because these are related to the $f$-values through the following expression, common for all types of transition:

\begin{equation}
f_{ij} = \frac{mc}{8{\pi}^2{e^2}}{\lambda^2_{ji}} \frac{{\omega}_j}{{\omega}_i} A_{ji}
 = 1.49 \times 10^{-16} \lambda^2_{ji} (\omega_j/\omega_i) A_{ji}
\end{equation}
where $m$ and $e$ are the electron mass and charge, respectively, $c$ the velocity of light,  $\lambda_{ji}$ the transition wavelength in \AA, and $\omega_i$ and $\omega_j$ the statistical weights of the lower ($i$) and upper ($j$) levels, respectively. Similarly,  $A_{ij}$  and $f_{ij}$ are related to the line strength $S$ (in atomic unit, 1 a.u. = 6.460$\times$10$^{-36}$ cm$^2$ esu$^2$) by the  following standard equations:

\begin{flushleft}
for the E1 transitions: 
\end{flushleft} 
\begin{equation}
A_{ji} = \frac{2.0261\times{10^{18}}}{{{\omega}_j}\lambda^3_{ji}} S \hspace*{1.0 cm} {\rm and} \hspace*{1.0 cm} 
f_{ij} = \frac{303.75}{\lambda_{ji}\omega_i} S, \\
\end{equation}
\begin{flushleft}
for the E2 transitions: 
\end{flushleft}
\begin{equation}
A_{ji} = \frac{1.1199\times{10^{18}}}{{{\omega}_j}\lambda^5_{ji}} S \hspace*{1.0 cm} {\rm and} \hspace*{1.0 cm}
f_{ij} = \frac{167.89}{\lambda^3_{ji}\omega_i} S, 
\end{equation}
\begin{flushleft}
and for the M1 transitions:  
\end{flushleft}
\begin{equation}
A_{ji} = \frac{2.6974\times{10^{13}}}{{{\omega}_j}\lambda^3_{ji}} S \hspace*{1.0 cm} {\rm and} \hspace*{1.0 cm}
f_{ij} = \frac{4.044\times{10^{-3}}}{\lambda_{ji}\omega_i} S. \\
\end{equation}

For many transitions the $f$-values are very small, i.e. the transitions are  weak and hence are likely not to make a  significant contribution in plasma modelling applications. For this reason, and to save space, only those transitions are listed in Tables 3 and 4 which have magnitudes $\ge$ 2\% of the largest $A$-value for any type. Similarly,  $A$-values for magnetic quadrupole (M2) and electric octupole (E3) transitions are not included in Tables 3 and 4. However, all $A$-values (in ASCII format) can be electronically obtained on request from K.Aggarwal@qub.ac.uk. Additionally, the level energies and the radiative transition parameters determined in the QRHF approximation can be found in the ADAMANT database  (http://www.adamant.tfai.vu.lt/database) at Vilnius University.

In the absence of other calculations it is difficult to assess the accuracy of our $A$-values. However, based on the accuracy of our calculated energy levels and our experience for a range of ions, we have confidence in our  data. Furthermore, in Tables 5 and 6 we list $\lambda$ and $f$-values for all {\em strong} ~E1 transitions of Co~IV and Ni~V, i.e. those having $f$ $\ge$ 0.10. Such comparatively strong transitions generally do not vary with varying amount of CI, and therefore we encourage other workers to perform independent calculations to assess the accuracy of our  data.

\section{Lifetimes}

The lifetime $\tau$ of a level $j$ is defined as:

\begin{equation}
{\tau}_j = \frac{1}{{\sum_{i}^{}} A_{ji}}
\end{equation}
where the sum is over all possible radiative decay channels $i<j$ and for all
calculated transition types. As for the $A$-values, to our knowledge there are  no existing  theoretical or experimental results  for $\tau$. Nevertheless, in Tables 1 and 2 we have included our values of $\tau$ for all levels and the calculations include contributions from {\em all} ~types of transition, i.e. E1, E2 and M1. These results will hopefully be useful for future comparisons and/or to further assess the accuracy of our calculations.  

\section{Conclusions}

 In this paper we have presented results for energy levels and  radiative rates for  transitions  of Co IV and Ni~V belonging to all 322 levels of the  3d$^6$,  3d$^5$4s and 3d$^5$4p configurations.  To our knowledge, there are no available results in the literature for $A$-values for comparison purposes. However,  energy levels are available, partly from measurements and mostly from analytical expressions and the other theoretical work for Ni~V. We have performed several sets of calculations with three independent atomic structure codes, namely {\sc grasp}, {\sc fac} and QRHF, with increasing amount of CI. It is noted that CI is very important for Cr-like ions in spite of Co and Ni being  moderately heavy. Our final results for all levels, obtained with the QRHF code, agree within 3\% with the NIST listings and there is no (significant) discrepancy in the level orderings.  It should  also be stressed that the level of CI included in the calculations was necessitated  by the availability of measured energy levels, otherwise results obtained would have been highly inaccurate.  We hope the present data  will be highly useful for the modelling of  plasmas.



\ack
The work at QUB has been partially funded  by AWE Aldermaston and in Romania from the Euratom research and training programme 2014-2018 under grant agreement No 633053 complementary research. The dissemination of results reflects only the authors' views and the Commission is not responsible for any use that may be made of the information it contains. Financial support under the Romanian Space Agency coordinated program STAR Project Number 33 is also acknowledged. P~B, R~K and R~K's research is funded by the European Social Fund under the Global Grant measure, project VP1-3.1-{\v S}MM-07-K-02-013. Finally, we thank an anonymous referee who made us aware of the recent calculations by Ong et al. \cite{ong1} for the energy levels of Ni~V.

\begin{appendix}

\def\thesection{} 

\section{Appendix A. Supplementary data}

Owing to space limitations, only parts of Tables 3 and 4  are presented here, the full tables being made available as supplemental material in conjunction with the electronic publication of this work. Supplementary data associated with this article can be found, in the online version, at doi:nn.nnnn/j.adt.2016.nn.nnn.

\end{appendix}



\clearpage
\newpage


\renewcommand{\baselinestretch}{1.0}
\footnotesize
\begin{longtable}{@{\extracolsep\fill}rlrrrr@{}}
\caption{Comparison of energies (in cm$^{-1}$) for the lowest 33 levels of Co IV.}
\\
Index  & Configuration/Level         &  NIST  & GRASP& FAC  & QRHF    \\
 \hline\\
\endfirsthead\\
\caption[]{(continued)}
Index  & Configuration/Level         &  NIST  & GRASP& FAC  & QRHF   \\
\hline\\
\endhead
  1  &  3d$^6\,\,\,^5_4$D$_4$            &      0.0  &     0  &     0  &      0    \\
  2  &  3d$^6\,\,\,^5_4$D$_3$            &    639.1  &   562  &   606  &    636    \\
  3  &  3d$^6\,\,\,^5_4$D$_2$            &   1077.7  &   956  &  1022  &   1076    \\
  4  &  3d$^6\,\,\,^5_4$D$_1$            &   1357.3  &  1209  &  1287  &   1358    \\
  5  &  3d$^6\,\,\,^5_4$D$_0$            &   1493.6  &  1333  &  1416  &   1496    \\
  6  &  3d$^6\,\,\,^3_4$P$_2$            &  22883.3  & 29081  & 24427  &  23613    \\
  7  &  3d$^6\,\,\,^3_4$H$_6$            &  23679.5  & 26414  & 27126  &  24357    \\
  8  &  3d$^6\,\,\,^3_4$H$_5$            &  24031.8  & 26753  & 27457  &  24717    \\
\hline  											       								   					      
\end{longtable}

                               
\begin{flushleft}                                                                               
                               
{\small
NIST: {\tt http://www.nist.gov/pml/data/asd.cfm} \\
GRASP: 14~732 level  calculations with the GRASP code \\
FAC: 76~138 level calculations with the FAC code \\
QRHF:  1~563~846   level calculations with the QRHF code \\                                                                               
}                                                                                               
                               
\end{flushleft} 

\renewcommand{\baselinestretch}{1.0}
\footnotesize
\begin{longtable}{@{\extracolsep\fill}rlll@{}}
\caption{ Comparison of theoretical energies (in cm$^{-1}$) for the 3d$^5$4s and 3d$^5$4p  levels of Co~IV.}
Index$^a$  & Configuration/Level                    & Present& Earlier$^b$    \\  \\
 \hline\\
\endfirsthead\\
\caption[]{(continued)}
Index$^a$  & Configuration/Level                    & Present& Earlier$^b$    \\  \\
\hline\\
\endhead 
 55  & 3d$^5(^4_3{\rm P})$4s~$^3$P$_0$              & 140554 & 141238    \\
 64  & 3d$^5(^2_5{\rm D})$4s~$^3$D$_1$              & 150602 & 151202    \\
 70  & 3d$^5(^4_3{\rm F})$4s~$^5$F$_1$              & 152090 & 152844    \\
 87  & 3d$^5(^4_3{\rm F})$4s~$^3$F$_3$              & 160219 & 160433    \\
 91  & 3d$^5(^2_5{\rm F})$4s~$^3$F$_2$              & 163283 & 163789    \\
 98  & 3d$^5(^2_5{\rm S})$4s~$^1$S$_0$              & 173942 & 173968    \\
 99  & 3d$^5(^2_3{\rm D})$4s~$^3$D$_1$              & 178332 & 178380    \\
117  & 3d$^5(^4_3{\rm P})$4p~$^5$D$_0^{\mathrm{o}}$ & 195340 & 195562.2\\
271  & 3d$^5(^2_5{\rm S})$4p~$^3$P$_0^{\mathrm{o}}$ & 233747 & 232730.5\\
301  & 3d$^5(^2_3{\rm P})$4p~$^3$P$_0^{\mathrm{o}}$ & 268904 & 267271.7\\
304  & 3d$^5(^2_3{\rm P})$4p~$^1$S$_0^{\mathrm{o}}$ & 272355 & 270879.8\\
319  & 3d$^5(^2_1{\rm D})$4p~$^3$P$_1^{\mathrm{o}}$ & 289735 & 287698.9\\
320  & 3d$^5(^2_1{\rm D})$4p~$^3$P$_0^{\mathrm{o}}$ & 290083 & 288081.3\\
322  & 3d$^5(^2_1{\rm D})$4p~$^1$P$_1^{\mathrm{o}}$ & 295845 & 292881.3\\
\\  \hline            								                	 
\end{longtable}   								   					       
			      							   					       

														       
\begin{flushleft}													       
{\small
$a$: see Table 1  \\
$b$: energies for the 3d$^5$4s levels are of van het Hof  et al. \cite{ams1} and for 3d$^5$4p are of Uylings  et al.  \cite{ams2} \\ 
}															       
\end{flushleft} 

\clearpage
\newpage

\renewcommand{\baselinestretch}{1.0}
\footnotesize
\begin{longtable}{@{\extracolsep\fill}rlll@{}}
\caption{ Comparison of theoretical energies (in cm$^{-1}$) for the 3d$^5$4s and 3d$^5$4p  levels of Ni~V.}
Index$^a$  & Configuration/Level                    & Present& Earlier$^b$    \\  \\
 \hline\\
\endfirsthead\\
\caption[]{(continued)}
Index$^a$  & Configuration/Level                    & Present& Earlier$^b$    \\  \\
\hline\\
\endhead 
 55  & 3d$^5(^4_3{\rm P})$4s~$^3$P$_0$              & 221924 & 221693    \\
 98  & 3d$^5(^2_5{\rm S})$4s~$^1$S$_0$              & 259591 & 258454    \\
153  & 3d$^5(^2_3{\rm P})$4s~$^3$P$_2$              & 297781 & 296114    \\
154  & 3d$^5(^2_3{\rm P})$4s~$^3$P$_1$              & 297821 & 296157    \\
155  & 3d$^5(^2_3{\rm P})$4s~$^3$P$_0$              & 297848 & 296187    \\
168  & 3d$^5(^2_3{\rm P})$4s~$^1$P$_0$              & 302675 & 300587    \\
116  & 3d$^5(^4_3{\rm P})$4p~$^5$D$_0^{\mathrm{o}}$ & 287705 & 287776.7  \\
304  & 3d$^5(^2_3{\rm P})$4p~$^1$S$_0^{\mathrm{o}}$ & 374539 & 373086.6  \\
319  & 3d$^5(^2_1{\rm D})$4p~$^3$P$_1^{\mathrm{o}}$ & 394868 & 393054.9  \\
321  & 3d$^5(^2_1{\rm D})$4p~$^3$P$_0^{\mathrm{o}}$ & 395370 & 393602.5  \\
322  & 3d$^5(^2_1{\rm D})$4p~$^1$P$_1^{\mathrm{o}}$ & 401752 & 399195.4  \\
\\  \hline            								                	 
\end{longtable}   								   					       
			      							   					       

														       
\begin{flushleft}													       
{\small
$a$: see Table 2  \\
$b$: energies for the 3d$^5$4s levels are of van het Hof  et al. \cite{ams1} and for 3d$^5$4p are of Uylings  et al.  \cite{ams2} \\ 
}															       
\end{flushleft} 

\clearpage
\newpage


\TableExplanation

\bigskip
\renewcommand{\arraystretch}{1.0}

\section*{Table 1.\label{tbl1te} Energies (in cm$^{-1}$) and  lifetimes ($\tau$, s) for the levels of Co~IV.}
\begin{tabular}{@{}p{1in}p{6in}@{}}
Index            & Level Index \\
Configuration    & The configuration to which the level belongs \\
Level             & The $LSJ$ designation of the level \\
NIST               & Energies compiled by NIST and available at  http://www.nist.gov/pml/data/asd.cfm  \\
QRHF           & Present energies from the QRHF code  with 1~563~846 level calculations \\
$\delta$E       & = 100\%(E$_{\mathrm{QRHF}}-E_{\mathrm{NIST}}$)/E$_{\mathrm{NIST}}$ is a relative percentage deviation  \\
$\tau$ (s)       & Lifetime of the level in s \\
$a{\pm}b$        &  $\equiv a{\times}$10$^{{\pm}b}$ \\
\end{tabular}
\label{tableII}

\bigskip
\renewcommand{\arraystretch}{1.0}

\section*{Table 2.\label{tbl2te}  Energies (in cm$^{-1}$) and  lifetimes ($\tau$, s) for the levels of Ni~V.}
\begin{tabular}{@{}p{1in}p{6in}@{}}
Index            & Level Index \\
Configuration    & The configuration to which the level belongs \\
Level             & The $LSJ$ designation of the level \\
NIST               & Energies compiled by NIST and available at  http://www.nist.gov/pml/data/asd.cfm  \\
QRHF           & Present energies from the QRHF code  with 1~563~846 level calculations \\
CI+MBPT        & Earlier calculations of Ong et al. \cite{ong1} \\
$\delta$E       & = 100\%(E$_{\mathrm{QRHF}}-E_{\mathrm{NIST}}$)/E$_{\mathrm{NIST}}$ is a relative percentage deviation  \\
$\tau$ (s)       & Lifetime of the level in s \\
$a{\pm}b$        &  $\equiv a{\times}$10$^{{\pm}b}$ \\
\end{tabular}
\label{tableII}


\bigskip
\section*{Table 3.\label{tbl3te}  Transition wavelengths in vacuum ($\lambda_{ij}$ in $\rm \AA$), radiative rates ($A_{ji}$ in s$^{-1}$),
 oscillator strengths ($f_{ij}$, dimensionless), and line strengths ($S$, in atomic units) for electric dipole (E1), and 
$A_{ji}$ for electric quadrupole (E2) and  magnetic dipole (M1) transitions of Co~IV.}
\begin{tabular}{@{}p{1in}p{6in}@{}}
$i$ and $j$         & The lower ($i$) and upper ($j$) levels of a transition as defined in Table 1.\\
$\lambda_{ij}$      & Transition wavelength (in ${\rm \AA}$) \\
$A^{E1}_{ji}$       & Radiative transition probability (in s$^{-1}$) for the E1 transitions \\
$f^{E1}_{ij}$       & Absorption oscillator strength (dimensionless) for the E1 transitions \\
S$^{E1}$            & Line strength in atomic unit (a.u.), 1 a.u. = 6.460$\times$10$^{-36}$ cm$^2$ esu$^2$ for the E1 transitions \\
$A^{E2}_{ji}$       & Radiative transition probability (in s$^{-1}$) for the E2 transitions \\
$A^{M1}_{ji}$       & Radiative transition probability (in s$^{-1}$) for the M1 transitions \\
$a{\pm}b$ &  $\equiv a\times{10^{{\pm}b}}$ \\
\end{tabular}
\label{ExplTable3}

\bigskip
\section*{Table 4.\label{tbl4te}  Transition wavelengths in vacuum ($\lambda_{ij}$ in $\rm \AA$), radiative rates ($A_{ji}$ in s$^{-1}$),
 oscillator strengths (f$_{ij}$, dimensionless), and line strengths ($S$, in atomic units) for electric dipole (E1), and 
$A_{ji}$ for electric quadrupole (E2) and magnetic dipole (M1) transitions of Ni~V.}
\begin{tabular}{@{}p{1in}p{6in}@{}}
$i$ and $j$         & The lower ($i$) and upper ($j$) levels of a transition as defined in Table 2.\\
$\lambda_{ij}$      & Transition wavelength (in ${\rm \AA}$) \\
$A^{E1}_{ji}$       & Radiative transition probability (in s$^{-1}$) for the E1 transitions \\
$f^{E1}_{ij}$       & Absorption oscillator strength (dimensionless) for the E1 transitions \\
$S^{E1}$            & Line strength in atomic unit (a.u.), 1 a.u. = 6.460$\times$10$^{-36}$ cm$^2$ esu$^2$ for the E1 transitions \\
$A^{E2}_{ji}$       & Radiative transition probability (in s$^{-1}$) for the E2 transitions \\
$A^{M1}_{ji}$       & Radiative transition probability (in s$^{-1}$) for the M1 transitions \\
$a{\pm}b$ &  $\equiv a\times{10^{{\pm}b}}$ \\
\end{tabular}
\label{ExplTable4}

\bigskip
\section*{Table 5.\label{tbl5te}  Radiative transition wavelengths  ($\lambda_{ij}$ in $\rm \AA$) and oscillator strengths ($f_{ij}$, dimensionless) for all strong E1 transitions of Co~IV.}
\begin{tabular}{@{}p{1in}p{6in}@{}}
$i$ and $j$         & The lower ($i$) and upper ($j$) levels of a transition as defined in Table 1.\\
$\lambda_{ij}$      & Transition wavelength (in ${\rm \AA}$) \\
$f_{ij}$                & Absorption oscillator strength (dimensionless) for the E1 transitions \\
\end{tabular}
\label{ExplTable5}

\bigskip
\section*{Table 6.\label{tbl6te}  Radiative transition wavelengths  ($\lambda_{ij}$ in $\rm \AA$) and oscillator strengths ($f_{ij}$, dimensionless) for all strong E1 transitions of Ni~V.}
\begin{tabular}{@{}p{1in}p{6in}@{}}
$i$ and $j$         & The lower ($i$) and upper ($j$) levels of a transition as defined in Table 1.\\
$\lambda_{ij}$      & Transition wavelength (in ${\rm \AA}$) \\
$f_{ij}$                & Absorption oscillator strength (dimensionless) for the E1 transitions \\
\end{tabular}
\label{ExplTable6}

\datatables 



\setlength{\LTleft}{0pt}
\setlength{\LTright}{0pt} 


\setlength{\tabcolsep}{0.5\tabcolsep}

\renewcommand{\arraystretch}{1.0}

\footnotesize 

\begin{longtable}{@{\extracolsep\fill}rllrrrr@{}}
\caption{Energies  (in cm$^{-1}$) and lifetimes ($\tau$, s) for the  levels  of  Co IV.  $a{\pm}b$  $\equiv a{\times}$10$^{{\pm}b}$. See page\ \pageref{tbl1te} for Explanation of Tables}
 Index$^c$ & Configuration & Level & NIST & QRHF & $\delta E \ $& $\tau$ (s)  \\
\hline\\
\endfirsthead\\
\caption[]{(continued)}
Index$^c$ & Configuration & Level & NIST & QRHF & $\delta E \ $& $\tau$ (s)  \\
\hline\\
\endhead
1  & 3d$^6$\,\,\, &   $^5_4$D$_4^{\mathrm{ }}$                   &  0.0       & 0      & 0     &	     \\
2  & 3d$^6$\,\,\, &   $^5_4$D$_3^{\mathrm{ }}$           	 &  639.1     & 636    & -0.49 & 1.130$+$02  \\
3  & 3d$^6$\,\,\, &   $^5_4$D$_2^{\mathrm{ }}$           	 &  1077.7    & 1076   & -0.16 & 1.828$+$02  \\
4  & 3d$^6$\,\,\, &   $^5_4$D$_1^{\mathrm{ }}$           	 &  1357.3    & 1358   & 0.05  & 4.777$+$02  \\
5  & 3d$^6$\,\,\, &   $^5_4$D$_0^{\mathrm{ }}$           	 &  1493.6    & 1496   & 0.16  & 2.383$+$03  \\
6  & 3d$^6$\,\,\, &   $^3_4$P$_2^{\mathrm{ }}$           	 &  22883.3   & 23613  & 3.19  & 9.484$-$01  \\
7  & 3d$^6$\,\,\, &   $^3_4$H$_6^{\mathrm{ }}$           	 &  23679.5   & 24357  & 2.86  & 8.668$+$04  \\
8  & 3d$^6$\,\,\, &   $^3_4$H$_5^{\mathrm{ }}$           	 &  24031.8   & 24717  & 2.85  & 8.037$+$02  \\
\end{longtable}
\begin{flushleft}													       
{\small
$c$: see text in section 2 for levels in bold faces  \\
	       
}															       
\end{flushleft} 

\clearpage
\newpage

\setlength{\tabcolsep}{0.5\tabcolsep}

\renewcommand{\arraystretch}{1.0}

\footnotesize 

\begin{longtable}{@{\extracolsep\fill}rllrrrrr@{}}
\caption{Energies  (in cm$^{-1}$) and lifetimes ($\tau$, s) for the  levels  of  Ni~V.   $a{\pm}b$  $\equiv a{\times}$10$^{{\pm}b}$.  See page\ \pageref{tbl2te} for Explanation of Tables}
 Index$^c$ & Configuration & Level & NIST & QRHF & CI+MBPT & $\delta E \ $& $\tau$ (s)  \\
\hline\\
\endfirsthead\\
\caption[]{(continued)}
Index$^c$ & Configuration & Level & NIST & QRHF & CI+MBPT & $\delta E \ $& $\tau$ (s)  \\
\hline\\
\endhead
1  & 3d$^6$          & $^5_4$D$_4$                     &     0.0  &	0    &         & 0.00	&	    \\   
2  & 3d$^6$          & $^5_4$D$_3$                     &   889.7  &   878    &         & -1.32  & 4.294$+01$\\
3  & 3d$^6$          & $^5_4$D$_2$                     &  1489.9  &  1476    &         & -0.93  & 7.295$+01$\\
4  & 3d$^6$          & $^5_4$D$_1$                     &  1871.5  &  1857    &         & -0.77  & 1.926$+02$\\
5  & 3d$^6$          & $^5_4$D$_0$                     &  2057.6  &  2044    &         & -0.66  & 9.673$+02$\\
6  & 3d$^6$          & $^3_4$P$_2$                     & 26153.0  & 26655    &         & 1.92	& 4.392$-01$\\
7  & 3d$^6$          & $^3_4$H$_6$                     & 27111.2  & 27905    &         & 2.93	& 5.357$+04$\\
8  & 3d$^6$          & $^3_4$H$_5$                     & 27578.2  & 28374    &         & 2.89	& 3.595$+02$\\
\end{longtable}
\begin{flushleft}													       
{\small
$c$: see text in section 2 for levels in bold faces  \\
	       
}															       
\end{flushleft} 

\clearpage

\newpage

\renewcommand{\arraystretch}{1.0}

\footnotesize 

\begin{longtable}{@{\extracolsep\fill}rrrrrrrr@{}}
\caption{Transition wavelengths in vacuum ($\lambda_{ij}$ in $\rm \AA$), radiative rates ($A_{ji}$ in s$^{-1}$),
 oscillator strengths ($f_{ij}$, dimensionless), and line strengths ($S$, in atomic units) for electric dipole (E1), and 
$A_{ji}$ for electric quadrupole (E2) and  magnetic dipole (M1) transitions of Co~IV.  See page\ \pageref{tbl3te} for Explanation of Tables}
$i$ & $j$ & $\lambda_{ij}$ & $A^{{\rm E1}}_{ji}$  & $f^{{\rm E1}}_{ij}$ & $S^{{\rm E1}}$ & $A^{{\rm E2}}_{ji}$  & $A^{{\rm M1}}_{ji}$  \\
\hline\\
\endfirsthead\\
\caption[]{(continued)}
$i$ & $j$ & $\lambda_{ij}$ & $A^{{\rm E1}}_{ji}$  & $f^{{\rm E1}}_{ij}$ & $S^{{\rm E1}}$ & $A^{{\rm E2}}_{ji}$  & $A^{{\rm M1}}_{ji}$ \\
\hline\\
\endhead
    1 &    2 &  1.572$+$04 &  0.000$+$00 &  0.000$+$00 &  0.000$+$00 &  0.000$+$00 &  8.850$-$03 \\
    1 &    7 &  4.106$+$02 &  0.000$+$00 &  0.000$+$00 &  0.000$+$00 &  1.150$-$05 &  0.000$+$00 \\
    1 &    8 &  4.046$+$02 &  0.000$+$00 &  0.000$+$00 &  0.000$+$00 &  0.000$+$00 &  2.570$-$05 \\
    1 &    9 &  4.004$+$02 &  0.000$+$00 &  0.000$+$00 &  0.000$+$00 &  0.000$+$00 &  2.440$-$02 \\
    1 &   12 &  3.824$+$02 &  0.000$+$00 &  0.000$+$00 &  0.000$+$00 &  0.000$+$00 &  9.740$-$01 \\
    1 &   13 &  3.775$+$02 &  0.000$+$00 &  0.000$+$00 &  0.000$+$00 &  0.000$+$00 &  9.540$-$02 \\
    1 &   15 &  3.352$+$02 &  0.000$+$00 &  0.000$+$00 &  0.000$+$00 &  0.000$+$00 &  9.470$-$04 \\
    1 &   16 &  3.290$+$02 &  0.000$+$00 &  0.000$+$00 &  0.000$+$00 &  0.000$+$00 &  2.220$-$02 \\
\end{longtable}

\newpage

\renewcommand{\arraystretch}{1.0}

\footnotesize 

\begin{longtable}{@{\extracolsep\fill}rrrrrrrrrr@{}}
\caption{Transition wavelengths in vacuum ($\lambda_{ij}$ in $\rm \AA$), radiative rates ($A_{ji}$ in s$^{-1}$),
 oscillator strengths ($f_{ij}$, dimensionless), and line strengths ($S$, in atomic units) for electric dipole (E1), and 
$A_{ji}$ for electric quadrupole (E2) and  magnetic dipole (M1) transitions of Ni~V. See page\ \pageref{tbl4te} for Explanation of Tables}
$i$ & $j$ & $\lambda_{ij}$ & $A^{{\rm E1}}_{ji}$  & $f^{{\rm E1}}_{ij}$ & $S^{{\rm E1}}$ & $A^{{\rm E2}}_{ji}$  & $A^{{\rm M1}}_{ji}$  \\
\hline\\
\endfirsthead\\
\caption[]{(continued)}
$i$ & $j$ & $\lambda_{ij}$ & $A^{{\rm E1}}_{ji}$  & $f^{{\rm E1}}_{ij}$ & $S^{{\rm E1}}$ & $A^{{\rm E2}}_{ji}$  & $A^{{\rm M1}}_{ji}$ \\
\hline\\
\endhead
    1 &    2 &  1.139$+$04 &  0.000$+$00 &  0.000$+$00 &  0.000$+$00 &  0.000$+$00 &  2.329$-$02 \\
    1 &    7 &  3.584$+$02 &  0.000$+$00 &  0.000$+$00 &  0.000$+$00 &  1.867$-$05 &  0.000$+$00 \\
    1 &    8 &  3.524$+$02 &  0.000$+$00 &  0.000$+$00 &  0.000$+$00 &  0.000$+$00 &  1.153$-$04 \\
    1 &    9 &  3.489$+$02 &  0.000$+$00 &  0.000$+$00 &  0.000$+$00 &  0.000$+$00 &  1.705$-$01 \\
    1 &   11 &  3.359$+$02 &  0.000$+$00 &  0.000$+$00 &  0.000$+$00 &  0.000$+$00 &  1.960$+$00 \\
    1 &   13 &  3.311$+$02 &  0.000$+$00 &  0.000$+$00 &  0.000$+$00 &  0.000$+$00 &  2.074$-$01 \\
    1 &   15 &  2.939$+$02 &  0.000$+$00 &  0.000$+$00 &  0.000$+$00 &  0.000$+$00 &  2.745$-$03 \\
    1 &   16 &  2.872$+$02 &  0.000$+$00 &  0.000$+$00 &  0.000$+$00 &  0.000$+$00 &  6.285$-$02 \\
\end{longtable}

\newpage

\renewcommand{\arraystretch}{1.0}

\footnotesize 

\begin{longtable}{@{\extracolsep\fill}rrrrrrrrrrrrrrrr@{}}
\caption{Radiative transition wavelengths  ($\lambda_{ij}$ in $\rm \AA$) and oscillator strengths ($f_{ij}$, dimensionless) for all strong E1 transitions of Co~IV. See page\ \pageref{tbl5te} for Explanation of Tables and Table 1 for definition of level indices.}
$i$ & $j$ & $\lambda_{ij}$ &  $f_{ij}$ &  $i$ & $j$ & $\lambda_{ij}$ &  $f_{ij}$ & $i$ & $j$ & 
$\lambda_{ij}$ &  $f_{ij}$ & $i$ & $j$ & $\lambda_{ij}$ &  $f_{ij}$  \\
\hline\\
\endfirsthead\\
\caption[]{(continued)}
$i$ & $j$ & $\lambda_{ij}$ &  $f_{ij}$ &  $i$ & $j$ & $\lambda_{ij}$ &  $f_{ij}$ & $i$ & $j$ & 
$\lambda_{ij}$ &  $f_{ij}$ & $i$ & $j$ & $\lambda_{ij}$ &  $f_{ij}$   \\
\hline\\
\endhead
 1& 118&  51.188& 0.108& 52& 129& 162.272& 0.173& 70& 224& 144.317& 0.138&  96& 257& 162.114& 0.138\\
 5& 124&  51.346& 0.100& 52& 136& 159.532& 0.327& 71& 186& 159.592& 0.100&  96& 264& 158.759& 0.154\\
11& 216&  51.536& 0.181& 52& 148& 149.186& 0.243& 71& 205& 150.731& 0.147&  96& 274& 146.990& 0.140\\
18& 212&  54.778& 0.107& 53& 137& 167.420& 0.140& 71& 206& 150.517& 0.297&  96& 276& 142.359& 0.253\\
23& 231&  55.457& 0.254& 53& 152& 157.305& 0.238& 71& 215& 148.112& 0.109&  97& 272& 155.333& 0.213\\
23& 287&  48.286& 0.102& 53& 168& 144.551& 0.117& 72& 192& 154.348& 0.274&  97& 273& 153.170& 0.370\\
24& 238&  55.355& 0.101& 54& 154& 156.999& 0.175& 72& 216& 148.599& 0.125&  98& 275& 156.371& 0.607\\
25& 276&  53.825& 0.114& 54& 158& 155.353& 0.134& 72& 217& 147.988& 0.102&  98& 287& 131.837& 0.168\\
\end{longtable}

\newpage

\renewcommand{\arraystretch}{1.0}

\footnotesize 

\begin{longtable}{@{\extracolsep\fill}rrrrrrrrrrrrrrrr@{}}
\caption{Radiative transition wavelengths  ($\lambda_{ij}$ in $\rm \AA$) and oscillator strengths ($f_{ij}$, dimensionless) for all strong E1 transitions of Ni~V. See page\ \pageref{tbl6te} for Explanation of Tables and Table 2 for definition of level indices.}
$i$ & $j$ & $\lambda_{ij}$ &  $f_{ij}$ &  $i$ & $j$ & $\lambda_{ij}$ &  $f_{ij}$ & $i$ & $j$ & 
$\lambda_{ij}$ &  $f_{ij}$ & $i$ & $j$ & $\lambda_{ij}$ &  $f_{ij}$  \\
\hline\\
\endfirsthead\\
\caption[]{(continued)}
$i$ & $j$ & $\lambda_{ij}$ &  $f_{ij}$ &  $i$ & $j$ & $\lambda_{ij}$ &  $f_{ij}$ & $i$ & $j$ & 
$\lambda_{ij}$ &  $f_{ij}$ & $i$ & $j$ & $\lambda_{ij}$ &  $f_{ij}$   \\
\hline\\
\endhead
    1 &   120 &      34.741 &       0.101 &   51 &   150 &     125.063 &       0.222 &   69 &	196 &	  127.422 &	  0.193 &   93 &   276 &     119.574 &       0.237   \\
   12 &   217 &      34.965 &       0.169 &   52 &   129 &     135.778 &       0.160 &   70 &	195 &	  128.503 &	  0.155 &   97 &   271 &     130.669 &       0.155   \\
   19 &   213 &      36.616 &       0.100 &   52 &   136 &     132.863 &       0.312 &   70 &	211 &	  124.779 &	  0.103 &   97 &   273 &     128.559 &       0.336   \\
   19 &   260 &      35.075 &       0.154 &   52 &   149 &     125.108 &       0.227 &   70 &	229 &	  121.221 &	  0.143 &   98 &   275 &     131.355 &       0.552   \\
   23 &   232 &      36.878 &       0.238 &   53 &   137 &     139.744 &       0.126 &   71 &	209 &	  126.192 &	  0.276 &   98 &   287 &     110.896 &       0.172   \\
   23 &   287 &      33.189 &       0.101 &   53 &   152 &     131.357 &       0.226 &   72 &	194 &	  129.979 &	  0.232 &   99 &   277 &     126.121 &       0.185   \\
   25 &   276 &      36.088 &       0.112 &   53 &   169 &     121.117 &       0.114 &   73 &	205 &	  128.212 &	  0.127 &   99 &   279 &     125.220 &       0.151   \\
   32 &   299 &      35.638 &       0.104 &   54 &   157 &     131.097 &       0.175 &   73 &	215 &	  126.389 &	  0.116 &   99 &   280 &     124.558 &       0.202   \\
\end{longtable}

\end{document}